# Quasiparticle interference of the Fermi arcs and surface-bulk connectivity of a Weyl semimetal

Hiroyuki Inoue[1]*, András Gyenis[1]*, Zhijun Wang[1], Jian Li[1], Seong Woo Oh[1], Shan Jiang[2], Ni Ni[2], B. Andrei Bernevig[1], Ali Yazdani[1]†

[1] *Joseph Henry Laboratories of Physics, Department of Physics, Princeton University, Princeton NJ 08540, USA*

[2] *Department of Physics and Astronomy and California NanoSystems Institute, University of California at Los Angeles, Los Angeles, CA 90095, USA*

*These authors contributed equally to this work

†Corresponding author. Email: yazdani@princeton.edu

**Weyl semimetals host topologically protected surface states, with arced Fermi surface contours that are predicted to propagate through the bulk when their momentum matches that of the surface projections of the bulk's Weyl nodes. We use spectroscopic mapping with a scanning tunneling microscope to visualize quasi-particle scattering and interference at the surface of the Weyl semimetal TaAs. Our measurements reveal 10 different scattering wave vectors, which can be understood and precisely reproduced with a theory that takes into account the shape, spin texture, and momentum-dependent propagation of the Fermi arc surface states into the bulk. Our findings provide evidence that Weyl nodes act as sinks for electron transport on the surface of these materials.**

Understanding the exotic properties of quasi-particles at the boundaries of topological phases of electronic matter is at the forefront of condensed matter physics research. Examples include helical Dirac fermions on the surface of time-reversal invariant topological insulators (*1*), which have been demonstrated to be immune to backscattering (*2-4*), or the emergent Majorana fermions at the edge of a topological superconductor (*5-7*). Topological properties of phases of matter are however not limited to gapped systems, and recent theoretical efforts have uncovered the possibility of topologically protected metallic phases. Topological Dirac semimetals (*8, 9*)—three-dimensional analogs of graphene with band crossings protected by crystalline symmetry—have been recently realized experimentally (*10-12*). Breaking of the inversion or time reversal symmetry splits the Dirac crossings of the band structure into Weyl points, which are singularities in the Berry curvature characterized by a Chern number or more colloquially monopole charge (*13, 14*). These semimetals host bulk quasiparticles that are chiral Weyl fermions. The topologically protected boundary modes of Weyl semimetals are surface states with a disjointed two-dimensional Fermi surface (*15, 16*). These so-called Fermi arcs connect surface projections of bulk Weyl points of opposing Chern numbers. A notable property of these Fermi arc states is that they can become delocalized into the bulk through the projected Weyl points. Ideally, in the absence of disorder, an electron on the Fermi arcs of one surface can sink through the bulk and appear on the arcs of the opposing surface.

Recent work has shown strong evidence that inversion symmetry breaking transition metal compounds (TaAs, TaP, NbAs, NbP) are Weyl semimetals (*17-21*). Angle resolved photoemission spectroscopy (ARPES) measurements have confirmed that the surface electronic structure of these compounds has a disconnected arc-like topology connecting the surface projection of 24 Weyl crossings in the bulk band structure. Here we used the scanning tunneling microscope (STM) to directly visualize the surface states of the Weyl semimetal TaAs and examine their scattering and quantum interference properties. Our experimental approach follows similar STM studies that showed that the spin texture (and time-reversal symmetry) protects surface states on topological insulators from backscattering (*2-4, 22*). In contrast to these previous studies, we find that the momentum-dependent delocalization of the Fermi arcs into the bulk, which is a unique property of these surface states, determines the coherent interference properties of these surfaces states. The surface-bulk connectivity examined here underlies several other novel electronic phenomena, such as non-local transport that can occur in Weyl semimetals (*23-27*). A recent magneto-transport study has observed quantum oscillations that may be associated with the nonlocal nature of transport with electron orbits traversing through bulk Weyl nodes and surface states (*28*).

To probe the properties of the Weyl semimetal's Fermi arc states, we have carried out STM studies (at 40K) of in situ cleaved surfaces of TaAs single crystals that show atomically ordered terraces and tunneling density of states spectra consistent with a semi-metal (Fig. 1). The analysis of the atomic step edge heights in STM topographs shows only one type of surface for the cleaved samples, which is likely terminated by As atoms [see below and (*29*)]. Although the surface atomic structure visualized in the STM images has four-fold symmetry, the underlying electronic structure of this compound only has $C_{2v}$ symmetry, which is evident from its overall crystal structure shown in Fig. 1A. A $C_{4v}$ non-symmorphic symmetry is broken by the surface. The anisotropic local shape of electronic signatures caused by native surface defects shown in STM topography (Fig. 1, B and D), as well as the scattering of surface quasiparticles around such defects over long length scales displayed in STM conductance maps (Fig. 1E) both show a clear $C_{2v}$ symmetry.

Detailed information about the electronic properties of the surface states can be obtained from measurements of the quasi-particle interference (QPI) patterns in large area STM conductance maps. The Fourier transform of such STM conductance maps identifies scattering wave vectors $q$, which connect $k_i$ and $k_f$ states on the contours of constant energy surface in momentum space (*30*). QPI measurements can be used not only to follow the evolution of the Fermi surface probed near the surface, but also to determine whether some scattering wave vectors are forbidden as a result of discrete symmetries (*2, 23, 24*). Information on quasiparticle lifetime can also be extracted from QPI; however, such effects do not typically result in momentum specific changes in the QPI, as we discuss here.

Experimentally, the detection of scattering wave vectors in QPI measurements is limited by instrumental resolution, which is determined by the stability of the instrument and the maximum averaging time possible during each map. In Fig. 2, we show QPI measurements on the TaAs surface obtained with a high-resolution, home-built STM instrument capable of averaging for up to 6 days while maintaining atomic register. The Fourier transform of these QPI measurements (Fig. 2), allows us to resolve the rich array of scattering wave vectors on this compound. Such high-resolution measurements are required to resolve the entire set of allowed scattering wave vectors in this compound. In Fig. 2, Q to S, we also show measurements of the QPI features as a function of energy along specific directions in $q$ space, displaying continuous dispersion of these features with energy, as is typical of QPI features. Recent measurements on a related Weyl compound (NbP) at lower resolution have yielded a subset of QPI signals (3 out of 10 wave vectors) reported here (*31*).

QPI measurements from surface states of most materials can be understood by starting from a model of contours of constant energy in momentum space, consistent with ARPES-measured band structure that can be used to calculate a joint density of states (JDOS) probability for scattering as a function of momentum difference $q$ (*32*). If the Femi surface is spin textured, as in the case of helical Dirac surface states or strongly spin-orbit coupled systems, the experimental results can be compared with the spin-dependent scattering probability (SSP) maps that also take into account the influence of relative spin orientations of the initial and final states on the QPI measurements (*2,29*). Following such an approach, we use density functional theory

(DFT) methods to calculate the surface states for TaAs. We reproduce both the shape (*17, 20, 21*) and spin texture (*33-35*) of the surface Fermi contours (Fig. 3A) recently measured with ARPES on this compound and calculate the SSP map near the chemical potential, which can then be compared to our experimental results at the same energy. In addition to the Fermi arcs, both ARPES measurements and the DFT calculations capturing them may include some features that are in fact caused by trivial surface states (*29*). This approach results in an SSP (Fig. 3B) that resembles the overall symmetry of the experimentally measured QPI pattern (Fig. 3F); however, it produces many more scattering wave vectors than seen experimentally (such as those in the middle of QPI zone highlighted in Fig. 3B missing in Fig. 3F). This approach or a related one proposed recently (*36*) for understanding the data on NbP (*31*) also fails to capture the QPI data on TaAs at other energies (*29*).

An accurate model of our experimental results over a wide energy range can be achieved if we consider not only the shape, density of states, and the spin texture of the Fermi arcs, but also their momentum dependent delocalization into the bulk of the sample. The key conceptual idea is that the QPI of the Fermi arcs is dominated by initial and final momentum states that do not strongly leak into the bulk. The degree of connectivity between Fermi arc surface states and the bulk is a property that depends on their momentum approaching the projection of Weyl points on the surface. It is also related to the atomic character of the Weyl nodes. Our DFT simulation of the TaAs electronic structure shows that the majority (~90%) of electronic states associated with the bulk Weyl nodes are based on Ta atomic orbitals (*29*). Consistent with this information, we also find from the DFT calculation that the electronic states close to the projected Weyl points on the surface arcs have a large component of such orbitals. This results in the slow decay of the Fermi arcs' spectral weight associated with Ta orbitals into the bulk, as compared to that associated with

As orbitals (Fig. 3C) (*29*). This picture suggests that in order to understand the QPI data on TaAs, we should project out the Fermi arc states that are associated with Ta and focus only on states with the As-orbital characteristics, which have the weakest connectivity to the bulk states and hence can interfere coherently to produce the QPI patterns. The Ta orbitals in the first layer only weakly contribute to the QPI signal: An electron that scatters from the As site to a Ta site in the surface layer is more likely to sink into the bulk Weyl states (Fig. 3C). The presence of topologically trivial surface states [likely the inner bowtie feature in Fig. 3, A and D; see also (*29*)] does not alter the projection of bulk Weyl points at the surface, and all surface states at Fermi level with these momenta would still be strongly delocalized into the bulk.

To confirm this idea, we consider a weighted Fermi arc contour for TaAs, in which we project the Fermi arc states onto the As orbitals at the top-most surface layer (Fig. 3D). This projection results in changes that are strongly dependent on the distance (in $k$ space) from the Weyl point. Taking into account the delocalization into the bulk, we find strong fading of the inner-most spoon-shaped Fermi arcs in the weighted Fermi surface, as compared to that shown in Fig. 3A. These short spoon-shaped sections of the Fermi arcs are close to their corresponding Weyl points, which are expected to act as sinks for electron propagation on the surface (because of their dominant Ta orbital character). In contrast, the prominent features of this weighted Fermi surface come from longer bowtie-shaped arcs at the Brillouin zone boundary, which lie far away from their corresponding Weyl point, with the least probability of sinking into the bulk (owing to their dominant As orbital character). An electron in the weighted Fermi arcs shown in Fig. 3D (or a similar one that also includes the As projection in the second layer; see the supplementary materials) is in a subset of states that do not efficiently propagate away into the bulk and remain near the surface.

Without any further computation, we can understand the various wave vectors seen in the QPI measurements by simply considering scattering around the weighted Fermi arc contours (Fig. 3D). Based on the length and orientation of the possible scattering wave vectors on the weighted Fermi surface, we identify ten different groups of scattering wave vectors, $Q_1$ to $Q_{10}$ that are seen experimentally. More detailed calculations of SSP (Fig. 3E) near the chemical potential based on these weighted Fermi arcs also compare favorably to the experimentally measured QPI at the same energy (Fig. 3F). In making this comparison, we note that although the spin texture of the Fermi arcs and the absence of backscattering between time reversed pairs of states play a role in the scattering data (a point to which we return below), the differences between JDOS and SSP for the Fermi arcs are relatively minor and not critical to understanding the QPI data [(*29*) and see below]. Contrasting results of the SSPs using weighted (Fig. 3E) and un-weighted (Fig. 3B) Fermi arcs with the experimental data (Fig. 3F) demonstrates that at some momenta the Fermi arc surface electrons have a strong probability of sinking into the bulk state and hence are not part of the QPI process. We further test this physical picture by a more exhaustive comparison of theoretical model calculations based on the weighted Fermi arcs and the experimental data over a wide range of energies (Fig. 4, A to L). As shown in this figure, our model SSP calculations for the weighted Fermi arc can reproduce the finer features of the large body of QPI data obtained in our studies. In addition, isolating the contribution from the As or Ta atomic sites to the QPI signal also can be used to confirm our theoretical identification of the major contribution of each atomic orbital to different sections of Fermi arcs [bowtie and spoon features dominated by As and Ta respectively (*29*)]. The agreement between theory and QPI measurements demonstrates the importance of accounting for delocalization of the TaAs surface states caused by the Weyl nature of its band structure.

We now return to the role of spin texture in determining the scattering properties of the Fermi arcs. Focusing on some of the finer features of the QPI data, in particular, scattering wave vectors $Q_3$ and $Q_9$, and their comparison to the theory, we can also resolve the subtle influence of spin texture on the STM data. This comparison (Figs. 4, M to P) demonstrates that some of the duplicate features in the JDOS maps caused by scattering between arc states that have opposing spins are suppressed in the SSP maps. Although almost at the limit of our experimental resolution, the correspondence between the finer predicted features in the SSP based on our model and those in the experimental QPI data qualitatively confirms that spin does play a role, albeit minor, in our QPI measurements. There are a few special wave vectors that are strictly prohibited owing to the presence of time reversal or mirror symmetry (*23, 24*). However, these signatures of protected scatterings are unfortunately obscured by many allowed overlapping wave vectors with similar lengths and are hard to resolve experimentally. Future experiments at higher resolution or on Weyl semimetal with simpler structures may better resolve this protection and other universal features of Fermi arcs that are predicted theoretically (*37, 38*).

Our work reveals that the absences of, or restriction on coherent scattering at certain wave vectors in both theory and experiments, are not a consequence of a symmetry related protection for the Fermi arc surface states, but rather follow from the topological connection between the surface and bulk states. This connection can also be seen in geometries in which the sample thickness is less than the bulk's scattering mean free path, where the top and bottom surface of a Weyl semimetal would be strongly linked through the Weyl points (*23, 26-28*).

**Acknowledgments:**
Work at Princeton was supported by Army Research Office-Multidisciplinary University Research Initiative (ARO-MURI) program W911NF-12-1-0461, Gordon and Betty Moore Foundation as part of the Emergent Phenomena in Quantum Systems (EPiQS) initiative (GBMF4530), by NSF-Materials Resarch Science and Engineering Centers (MRSEC) programs through the Princeton Center for Complex Materials DMR-1420541, NSF-DMR-1104612, NSF CAREER DMR-0952428, Packard Foundation, and Keck Foundation. This project was also made possible through use of the facilities at Princeton Nanoscale Microscopy Laboratory supported by grants through ARO-W911NF-1-0262, ONR-N00014-14-1-0330, ONR-N00014-13-10661, U.S.Department of Energy-Basic Energy Sciences (DOE-BES), Defense Advanced Research Projects Agency – U.S.Space and Naval Warfare Systems Command (DARPA-SPWAR) Meso program N6601-11-1-4110, LPS and ARO-W911NF-1-0606, and Eric and Wendy Schmidt Transformative Technology Fund at Princeton. Work at University of California-Los Angeles was supported by the DOE-BES (DE-SC0011978).

**Fig. 1. STM topography and *dI/dV* spectroscopy of Weyl semimetal TaAs**. (**A**) Illustration of the crystal structure of TaAs with (001) As termination. One unit cell contains four layers of As and Ta. The lattice parameters are $a = b = 3.43$ Å and $c = 11.64$ Å. (**B**) STM topographic image ($V_{bias} = 500$ mV, $I_{setpoint} = 100$ pA) of the cleaved (001) surface of TaAs. Magnified views on some of the pronounced impurities (right panels) show apparent $C_{2v}$ symmetric deformation of the electronic structure. (**C**) Spatial variation of the differential conductance measurements along a line of 30 Å, shown as a yellow line in (B), was measured with lock-in techniques employing 3 mV excitations at 707 Hz. The orange curve shows the spatially averaged *dI/dV* spectra. An atomically periodic modulation of the spectra is visible. (**D**) Topographic image and (**E**) conductance map at 120 meV ($V_{bias} = -340$ mV, $I_{setpoint} = 80$ pA) on the same area, where the $C_{2v}$ symmetric nature of the surface electronic states is clearly visible.

**Fig. 2. Quasi-particle interference of TaAs surface states.** (**A** to **D**), (**I** to **L**) Spatially resolved *dI/dV* conductance maps at different energies obtained on the area shown in Fig. 1B ($V_{bias} = 240$ meV, $I_{setpoint} = 120$ pA). (**E** to **H**) (**M** to **P**) Symmetrized and drift corrected Fourier transforms of the *dI/dV* maps (QPI maps). Red dot indicates the $(2\pi, 2\pi)$ point in the reciprocal space in the unit of $1/a$; and PSD denotes power spectral density. In the colorbar *m*, corresponds to the mean value of the map and $\sigma$ to the standard deviation. (**Q** to **S**) Energy-momentum structure of the *dI/dV* maps ($V_{bias} = 600$ meV, $I_{setpoint} = 80$ pA) along the high symmetry directions shown in (**M**).

**Fig. 3**. **Fermi arcs and quasi-particle interference.** (**A**) DFT-calculated Fermi arc contour of constant energy in first Brillouin zone (BZ) at + 40 meV calculated projecting the DFT calculated spectral density to the top unit cell. Green dots indicate the projected positions of the

Weyl-nodes and arrows show the direction of the spin on the Fermi arcs. The combination of time-reversal symmetry and $C_{2v}$ symmetry implies vanishing out-of-plane component of the spin. FS, Fermi Surface. (**B**) SSP derived from (A) marked by regions of strong scattering in the middle of the BZ which is missing in the QPI data. (**C**) The integrated spectral density over the full BZ for As and Ta separately as a function of layer index shows the fast decay of As orbital states. (**D**) Weighted Fermi surface (WFS) calculated by projecting the electronic states only to the top-most As layer. The $Q$ vectors indicate the scattering wave vectors expected. (**E**) SSP based on (D) with red boxes showing the scattering wave vectors mapped in (F). (**F**) QPI map at 40 meV (same as in Fig. 2G) and experimentally observed groups of $Q$ vectors. FT-STS, Fourier transform-scanning tunneling spectroscopy.

**Fig. 4. Comparison of the FT-STS maps with the DFT calculation at various energies**. (**A** to **C**), (**G** to **I**) QPI maps (same as in Fig. 2) and (**D** to **F**), (**J** to **L**) the corresponding SSP derived from the projection of the spectral density to the topmost As. The obtained data and the calculations are in agreement over a wide energy range. (**M** to **O**) Enlargement of the $Q_9$ and $Q_3$ vectors in the calculated JDOS, SSP and in the QPI data at different energies. Whereas strong triple-arc structures are visible in JDOS, SSP shows a reduced number of arcs, which is consistent with the single arcs in the QPI data.

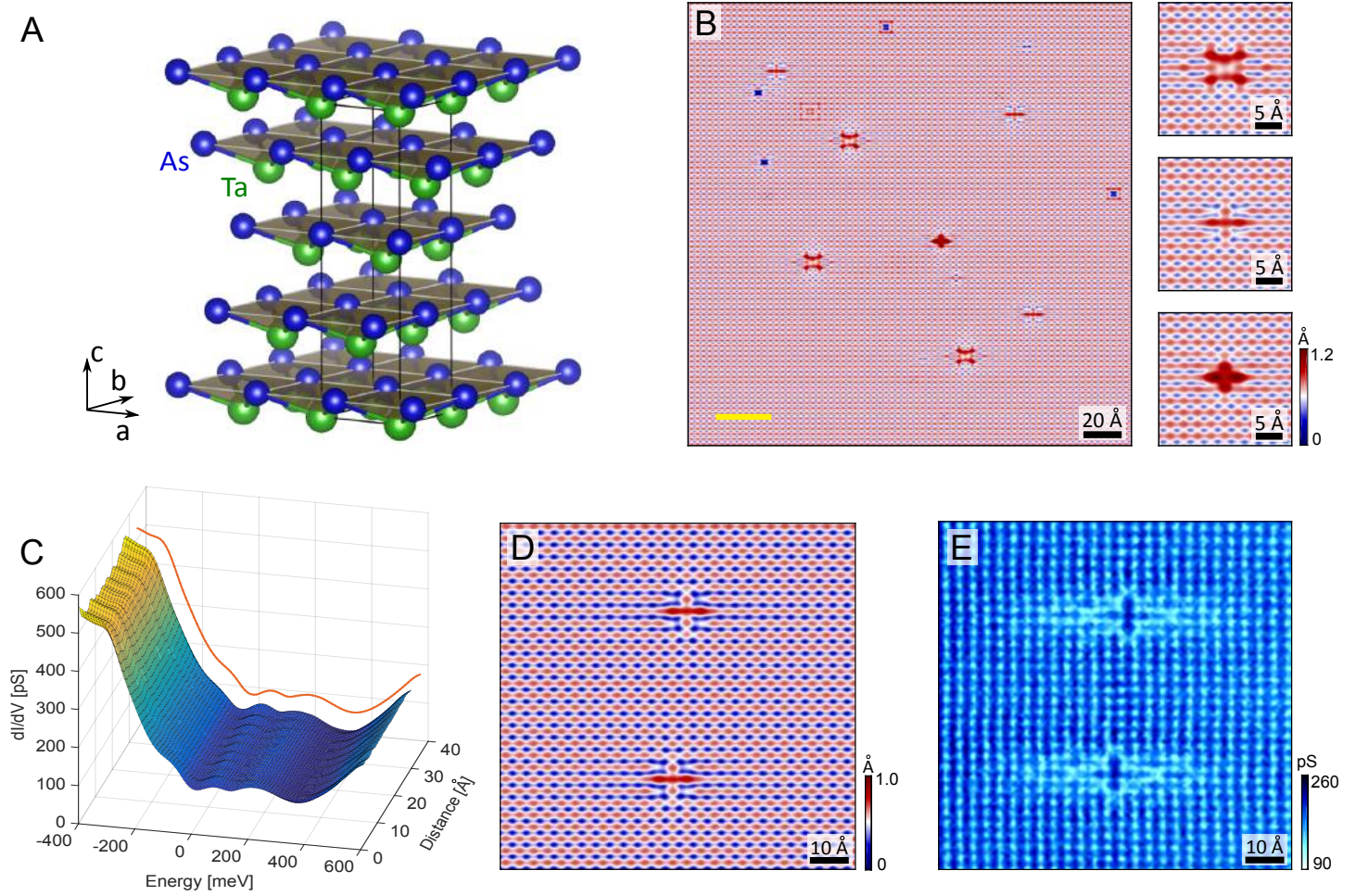

# Figure 2

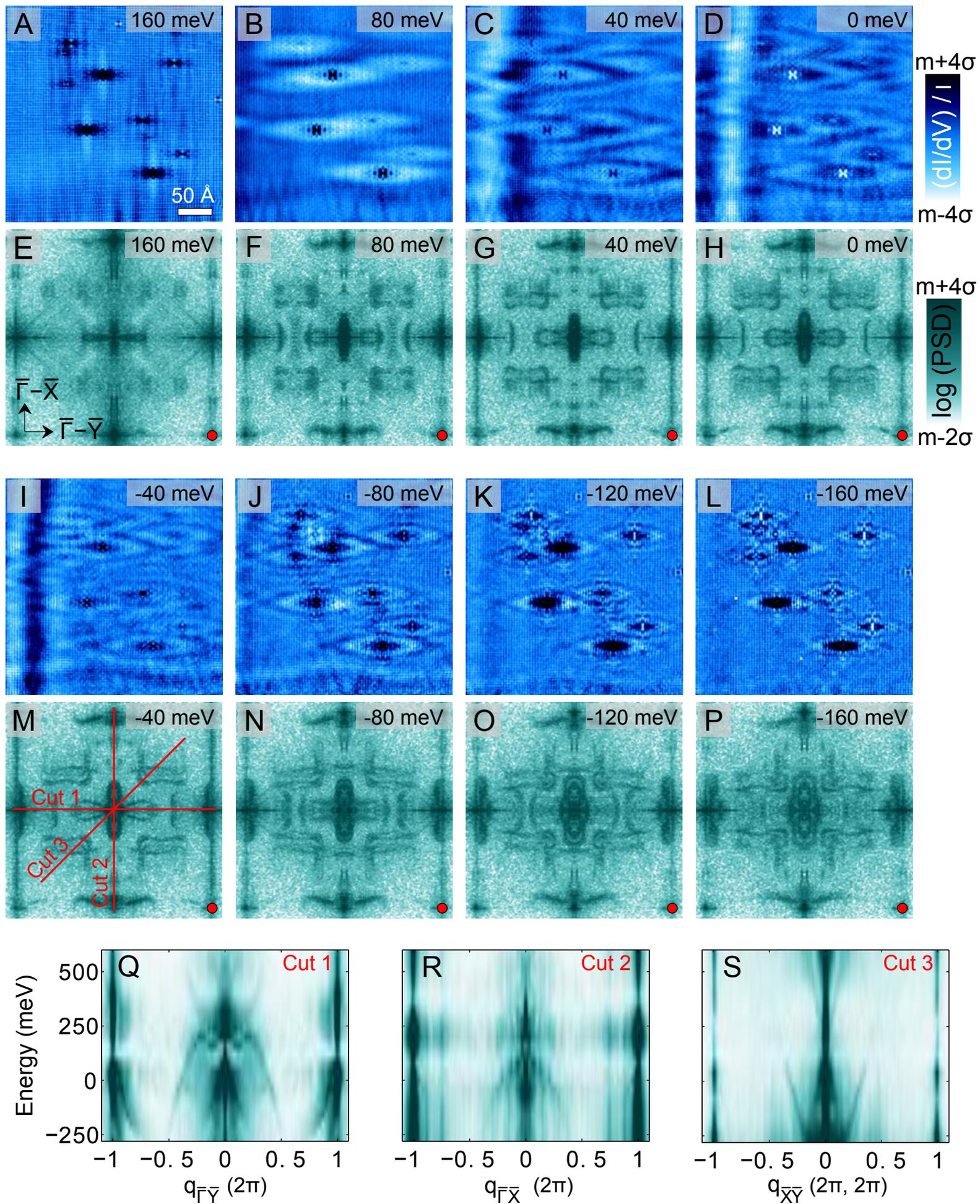

Figure 3

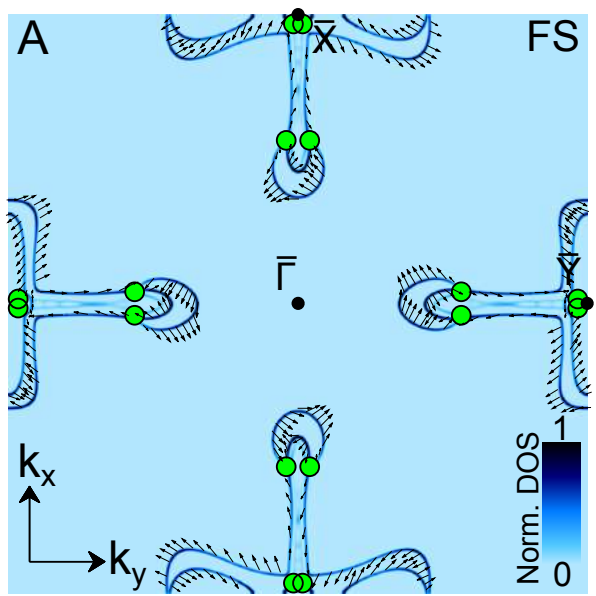
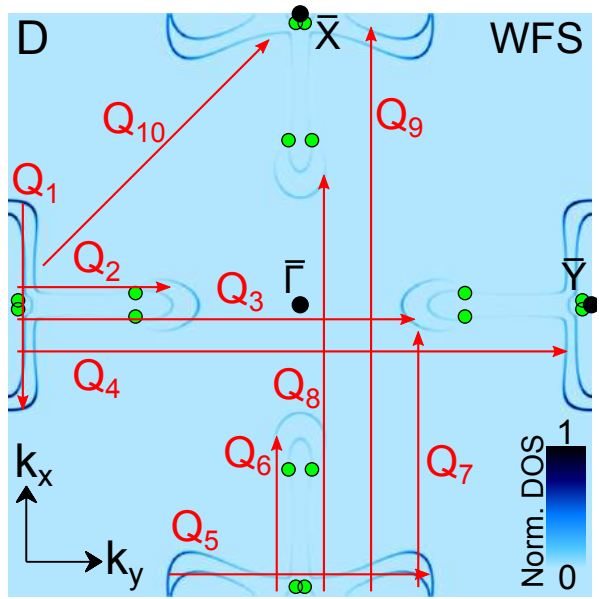
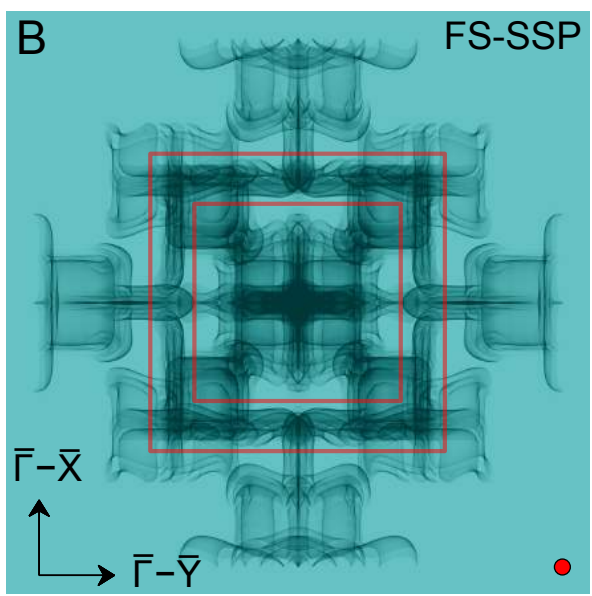
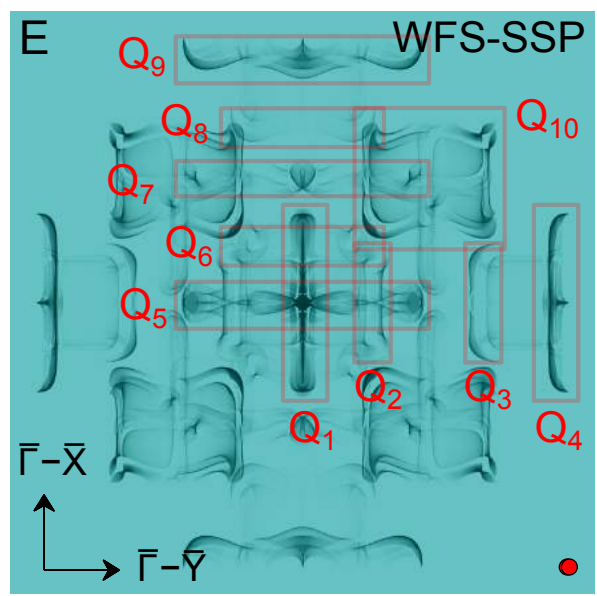
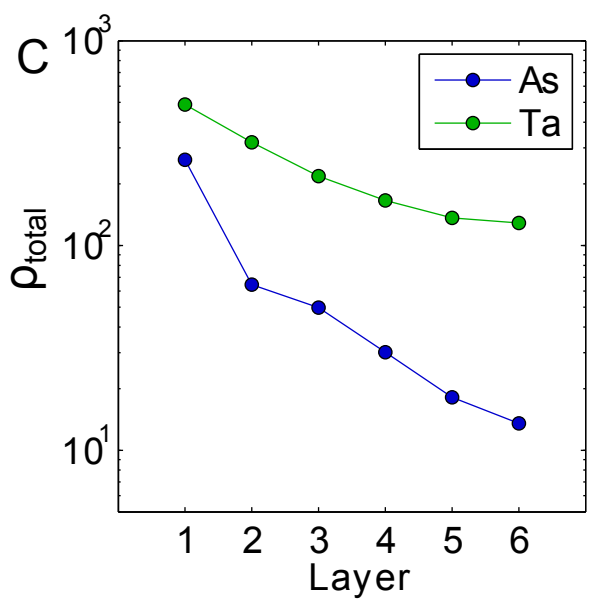
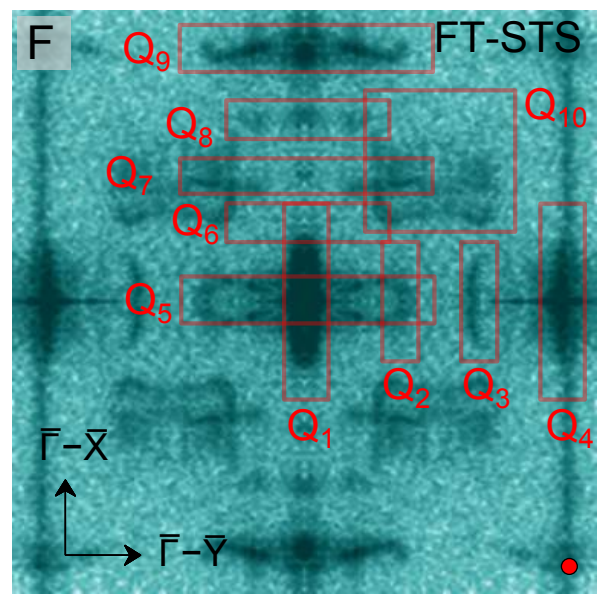

# Figure 4

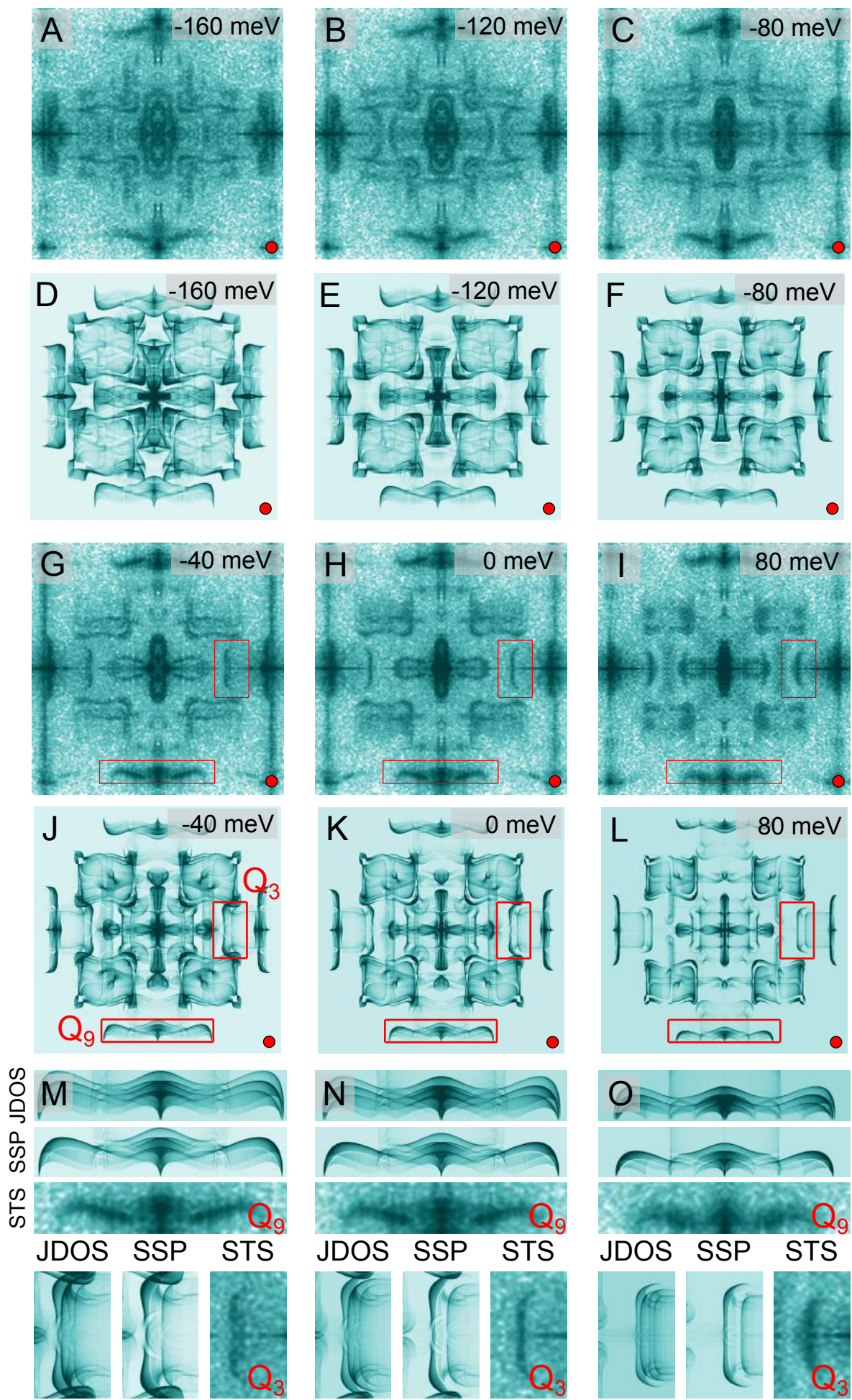

# Supplementary Materials for "Quasiparticle interference of the Fermi arcs and surface-bulk connectivity of a Weyl semimetal"

Hiroyuki Inoue[1]*, András Gyenis[1]*, Zhijun Wang[1], Jian Li[1], Seong Woo Oh[1], Shan Jiang[2], Ni Ni[2], B. Andrei Bernevig[1], Ali Yazdani[1]†

[1] *Joseph Henry Laboratories of Physics, Department of Physics, Princeton University, Princeton NJ 08540, USA*

[2] *Department of Physics and Astronomy and California NanoSystems Institute, University of California at Los Angeles, Los Angeles, CA 90095, USA*

*These authors contributed equally to this work

†Corresponding author. Email: yazdani@princeton.edu

## Materials and Methods

Sample growth

We grew TaAs single crystals (Fig. S1A) out of Ga flux with the ratio of Ta:As:Ga = 1:1:98. To increase the solubility of Ta in Ga, we used 80% of Ta in the thin foil form and 20% of Ta in the powder form as the starting material. We packed all starting materials in an alumina crucible, together with a catch crucible filled with quartz wool, then sealed them inside a quartz tube under 1/3 ATM Argon gas. The ampoule was heated up to 1100°C, held for 10 hours, cooled down to 1000°C using 3°C/h and held for 4 hours, then cooled down to 700°C at 1°C/h. We then took the ampoule out and spin off the Ga flux at ~80°C. Ga flux was removed from the single crystals using concentrated hydrochloride acid.

Surface termination

We carried out a large-scale topographic mapping (~ 900 Å x 900 Å) in order to investigate the (001) surface terminations. Fig. S1B reveals multiple terraces as a result of the cleaving procedure. The line cut along the green line (Fig. S1C) shows seven terraces (terrace index 1 to 7) with the height of each terrace of 2.72 Å (Fig. S1D), which is in good agreement with the nominal height difference between As layers (2.91 Å). The presence of equal, single step heights demonstrates that we observe only one kind of termination in the field of view.

As discussed in the main text, the theoretical calculation of the surface band structure for As termination matches remarkably well the observed interference patterns, while the calculated band structure for Ta termination is significantly different (Fig. S1E). Such differences between the Fermi surfaces of two different terminations can be attributed to, e.g., dangling-bond induced strong Rashba splitting and strong surface band bending (*35*).

As a consequence the Fermi surface on the Ta-terminated surface contains many more contributions from trivial surface bands than that on the As-terminated surface. Therefore, we conclude that the experimental observations are most likely associated with the As-terminated surface.

Details of DFT calculation

Our simulations consist of two major steps: extracting Fermi surface information (assuming various Fermi energies) from a full density-functional theory (DFT) calculation, and computing correlation functions such as joint density of states (JDOS) and spin scattering probability (SSP) by using the information obtained in the first step.

Explicitly, we first perform electronic structure calculations based on DFT as implemented in the Vienna ab initio simulation package (*39*), and use the core-electron projector augmented wave basis sets (*40*) with the generalized-gradient method (*41*). Spin-orbital coupling (SOC) is included self-consistently. The cutoff energy for wave-function expansion is 300 eV. Experimental lattice parameters are used to construct a slab model of nine surface unit cells in thickness. We use in-plane k-point grids of size $12 \times 12$ for the charge self-consistent calculations, and size $1000 \times 1000$ for the Fermi surface calculations.

The Fermi surfaces are calculated in terms of the total spectral density $\rho_0(\boldsymbol{k}) = \text{Tr}A(\boldsymbol{k})$, as well as the spin density $\rho_i(\boldsymbol{k}) = \text{Tr}[\sigma_i A(\boldsymbol{k})]$ ($i = 1, 2, 3$), with $A(\boldsymbol{k})$ being the spectral function matrix and $\sigma_{1,2,3}$ being the Pauli matrices for spin. The spectral function matrix $A(\boldsymbol{k})$ can be constructed from the Bloch eigenstates $\{\Psi_n(\boldsymbol{k})\}$ ($n$ is the band index), obtained from DFT calculations, for a chosen energy $E$, by taking its standard definition

$$A(\boldsymbol{k}, E) = \sum_n \left(-\frac{1}{\pi}\right) \text{Im}[\frac{1}{E - E_n(\boldsymbol{k}) + i\eta}] \Psi_n(\boldsymbol{k}) \Psi_n^\dagger(\boldsymbol{k}).$$

Here $E_n(\boldsymbol{k})$ is the energy of the n-th Bloch band, the eigenstate $\Psi_n(\boldsymbol{k})$ is a column vector, and $\eta$ is a small number typically of 3 meV.

Based on the spectral density $\rho_0(\boldsymbol{k})$ and the spin density $\rho_{1,2,3}(\boldsymbol{k})$ obtained from DFT calculations, the JDOS and SSP are calculated, respectively, as

$$J_0(\boldsymbol{q}) = \sum_{\boldsymbol{k}} \rho_0(\boldsymbol{k}) \rho_0(\boldsymbol{k} + \boldsymbol{q}),$$
$$J_s(\boldsymbol{q}) = \frac{1}{2} \sum_{\boldsymbol{k}} \sum_{i=0,1,2,3} \rho_i(\boldsymbol{k}) \rho_i(\boldsymbol{k} + \boldsymbol{q}).$$

The formula for SSP above is equivalent to one commonly used in the literature (*2*) for pure state limit where $\rho_0(\boldsymbol{k}) = 1$ and the sum of $\rho_i(\boldsymbol{k}) \rho_i(\boldsymbol{k} + \boldsymbol{q})$ over $i$=1, 2, 3 gives precisely $\cos(\theta)$ with $\theta$ the relative angle between the polarizations of the two states. The same equivalence holds also for mixed states by considering reduced density matrices (*37*). Here, in order to account for the fact that STM measurements render a higher resolution than the atomic lattice spacing, it is necessary to include extended Brillouin zones (BZs) in the above calculations. Phenomenologically, we maintain $\rho$'s in the first BZ as obtained from DFT calculations, and assume $\rho$'s in the extended BZs to be identically zero. The resulting JDOS and SSP agree well with experimental data in a large energy range.

**Supplementary Text**

Details of the Fermi arcs and possibility of trivial states:

The DFT calculations shown in Figure 3A match both the shape and spin texture of the surface states that have been measured on the surface of TaAs in recent ARPES measurements (*19-21*). However, there is still some ambiguity as to whether some of the features seen in the DFT and ARPES are due to topological or trivial surface states. In principle, pairs of trivial surface states can be present at the surface of a Weyl semimetal, without changing the bulk's topological classification. The only requirement is that a closed loop in the BZ that includes a single Weyl point possesses an odd number of Fermi-level crossings due to a single branch of Fermi arc that connects the conduction and valence bands. Figure S2A depicts a zoom-in view of the surface band structure at the Fermi level obtained by the DFT calculation where we choose two loops: a circle and square which encircle two and one Weyl points, respectively. Note that two bulk Weyl points are projected on top of each other in the circle. Figures S2C and D show the band structures along these contours. More detailed examination of the band crossings relative to the bulk conduction and valance bands (shaded gray regions in Figs. S2C and D) and the chemical potential can be used to determine which of the Fermi surface features are topological or trivial. The result of this analysis is schematically displayed in Fig. S2B (see the figure caption for the details). We note that the presence of trivial surface states does not alter the fact that the electronic states at the surface that hybridized with those at the Weyl point will sink into the bulk and will not participate in the QPI that we detect with the STM. The weighted Fermi surface of Fig. 3D, takes into account this connection regardless of the presence of topological trivial surface states.

Comparison of JDOS and SSP at various energies

Fig. S3 shows the calculated JDOS and SSP based on the weighted Fermi surface of the top-most As layer. The general shape and features of both are similar with only subtle differences between them. As discussed in the main text, taking into account of the spin-texture of the Fermi arcs only reduces some of the fine features seen in JDOS.

Comparison of SSPs of quadruple layer with that of top-most As layer at various energies

Fig. S4 demonstrates that SSPs at various energies based on the projection of the spectral density to the top quadruple layer exhibits far more features than those attributed to the top-most As layer. The obtained QPI data agrees well with the latter as also argued in the main text in details. Such excellent agreement exemplifies that the momentum-dependent delocalization of the surface Fermi arcs manifests at the wide energy range.

Weight distribution of Ta and As contents of bulk Weyl-node states

In Fig. S5B, we show a bulk band structure calculation, which identifies the positions of the Weyl nodes in the bulk Brillouin zone and the weight distribution of the Weyl node states on different atomic orbitals along the line cut through Weyl point 1 (W1) and 2 (W2) (Fig. S5A). Our results agree with previous studies of the positions of the Weyl nodes. Furthermore we find that the bulk Weyl node states are dominantly contributed by Ta 5d orbitals. The ratios of the two weights (Ta-5d vs. As-4p) at W1 and W2 are about 6:1 and 14:1, respectively.

Layer-depth dependence of weight distribution of projected surface states

We further consider $\rho_0$ and $\rho_{1,2,3}$ projected to layers of various thickness or compositions, e.g., a quadruple/double/single layer, and/or separate As/Ta atoms (a half layer) (Fig. S6A). In Fig. S6B to S6I, we show $\rho_0$ projected to As and Ta atoms, separately, in the top-most quadruple layer. We clearly see that in general different parts (features) of the Fermi surface may have different weight distributions on these atoms, and they penetrate with different length scale into the bulk of the slab. The correlation between the weight distribution and the penetration length is illustrated in Fig. S6B-S6I, where a significantly faster decay of total spectral density on As atoms than that on Ta atoms can be clearly seen. This is further confirmed by comparing the total spectral weight for each atom as a function of layer depth in logarithmic scale (Fig. 3C).

Atomic-site-selective QPI map

As an additional analysis, we describe an atomic-site-selective QPI analysis, which is generated by isolating the QPI signal in the *dI/dV* conductance maps from the As or Ta atomic sites. Following the crystal structure (Fig. 1A) and the identification of the surface termination of our system, we identify the maxima (minima) in Fig. 1B as As (Ta) sites. This provides a method to isolate the contribution of each atomic orbital to the QPI maps. Taking an experimentally measured conductance map at E = 40 meV (Fig. S7A), we generate the As-site (Ta-site) contribution to the conductance map as shown in Fig. S7B (Fig. S7C). The Fourier transform of each map is shown in Fig. S7D to S7F. This analysis, which allows the decomposition of Q maps, shows that the QPI features are strongest on the As atoms corresponds to the scattering among bowties features in the Fermi arcs (Fig. S7G), while the Ta-site QPI map mainly originates from scattering among spoons and bowties features on the arcs (Fig. S7H). This decomposition supports the theoretical identification of the majority contribution to each feature of Fermi arcs from each orbital. The weakness of the spoon-shape features in the weight Fermi surface that matches the experimental QPI data is due to their Ta orbital character, which as we discuss in the text leaks strongly into the bulk. Fig. S8 shows sets of QPI data sets and their As-site and Ta-site isolated contributions at various energies.

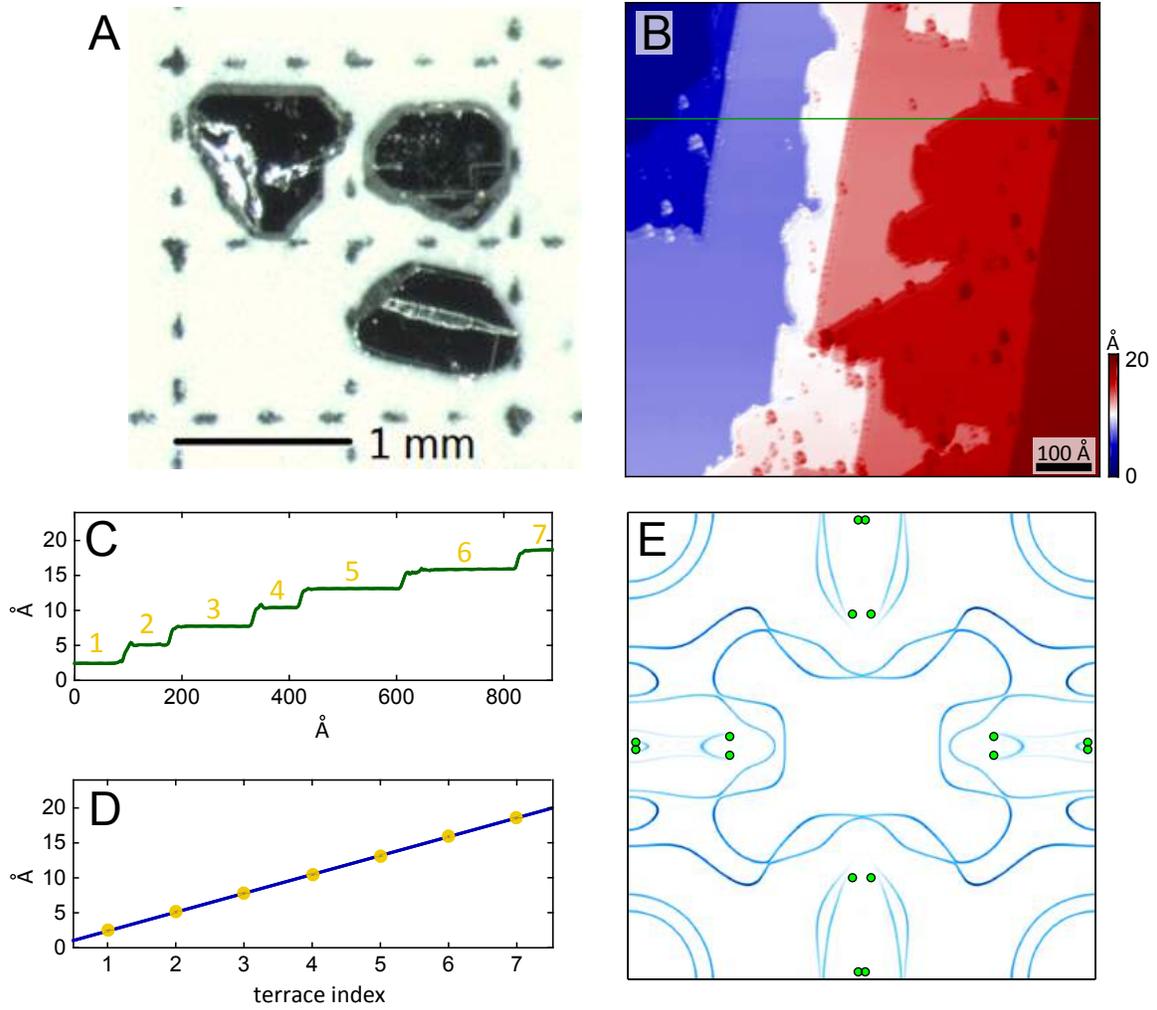

**Fig. S1.** (**A**) Picture of TaAs crystals taken by an optical microscope. (**B**) STM topographic mage ($V_{bias}$ = + 500 mV, $I_{setpoint}$ = 20 pA) of the cleaved (001) surface. (**C**) Line cut of the topographic profile along the green line in (**B**). In yellow numbers, terrace indexes 1 to 7 are assigned for the observed terraces. (**D**) With the terrace index as the horizontal axis, the average height of each terrace is plotted showing a step height of 2.71A. (**E**) Fermi surface projected to the surface layers with Ta termination.

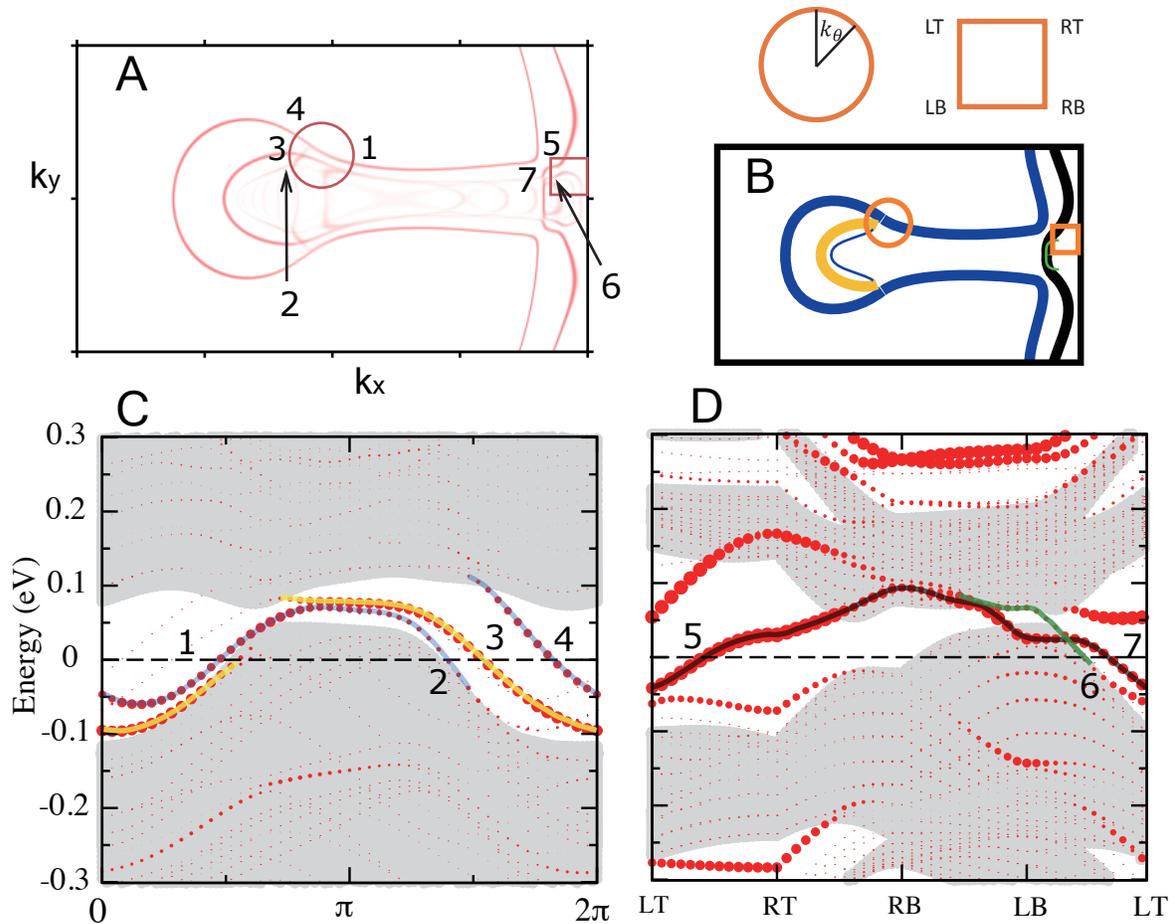

**Fig. S2.** (**A**) A portion of the surface FS projected to the outmost surface unit cell with As-termination. The topological nature of the equal energy contours, schematically indicated in (**B**), can be identified from the band structure (**C** and **D**) along two closed momentum-space trajectories each enclosing a Weyl-point projection (a spoon-end or a bowtie-end). The band structure plotted in (**C**) shows two topologically nontrivial surface bands (blue and yellow lines) that connect the conduction and valence bulk bands (gray shades) and are localized at the As termination (indicated by the size of the red dots). The two nontrivial bands are owing to the two Weyl points of the same chirality enclosed by the k-trajectory. The blue band gives rise to three crossing points (1, 2 and 4) at the Fermi energy, corresponding to the outer spoon (point 4), the outer bowtie (point 1), and a very weak feature (point 2) due to hybridization with bulk states. The band structure plotted in (**D**) shows one topologically nontrivial (green line) and one trivial (black line) surface band, owing to the single Weyl point enclosed by the k-trajectory. The trivial surface band gives rise to the inner bowtie, whereas the nontrivial one leads to a small Fermi arc that is probably not seen in experiments.

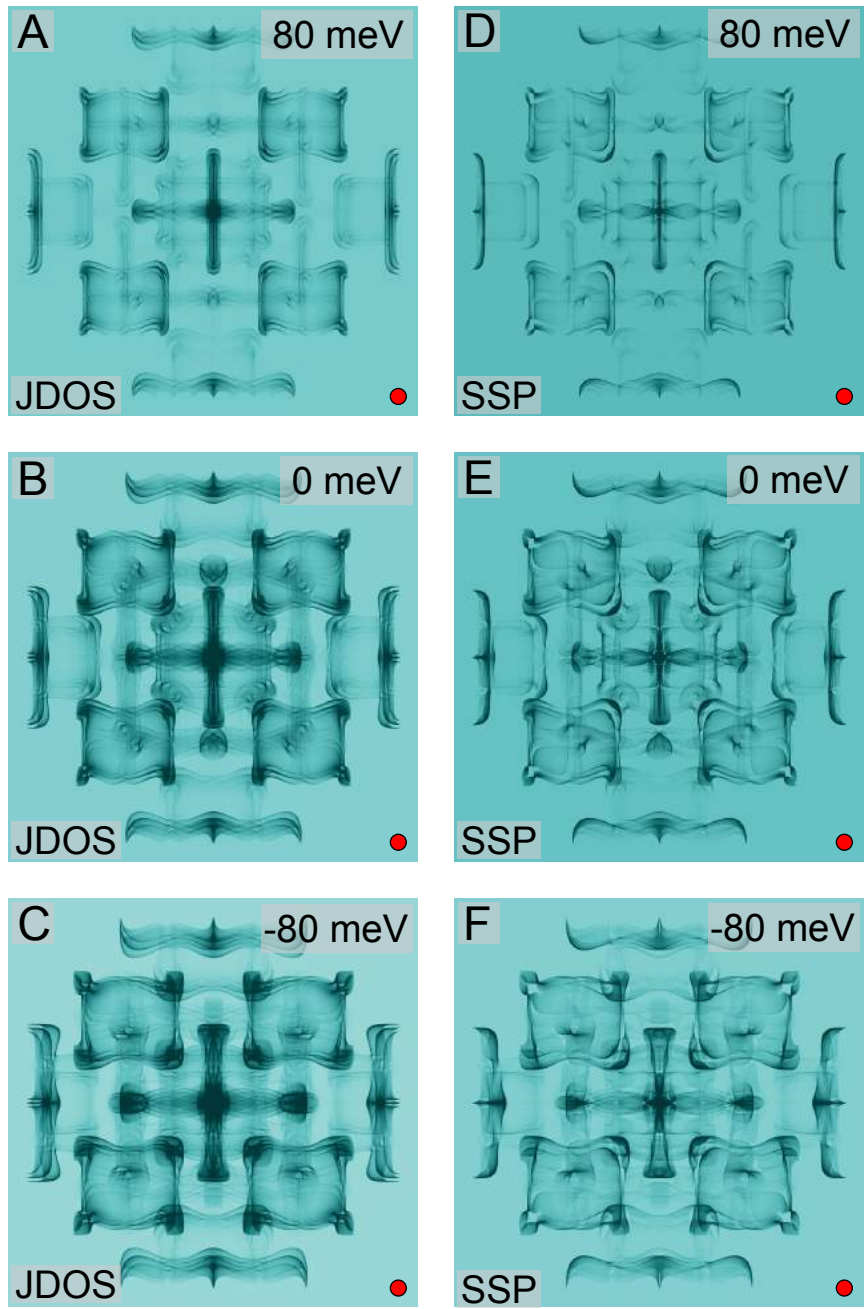

**Fig. S3.** Comparison of JDOS (**A-C**) and SSP (**D-F**) derived from the weighted Fermi surface of the top-most As layer at various energies.

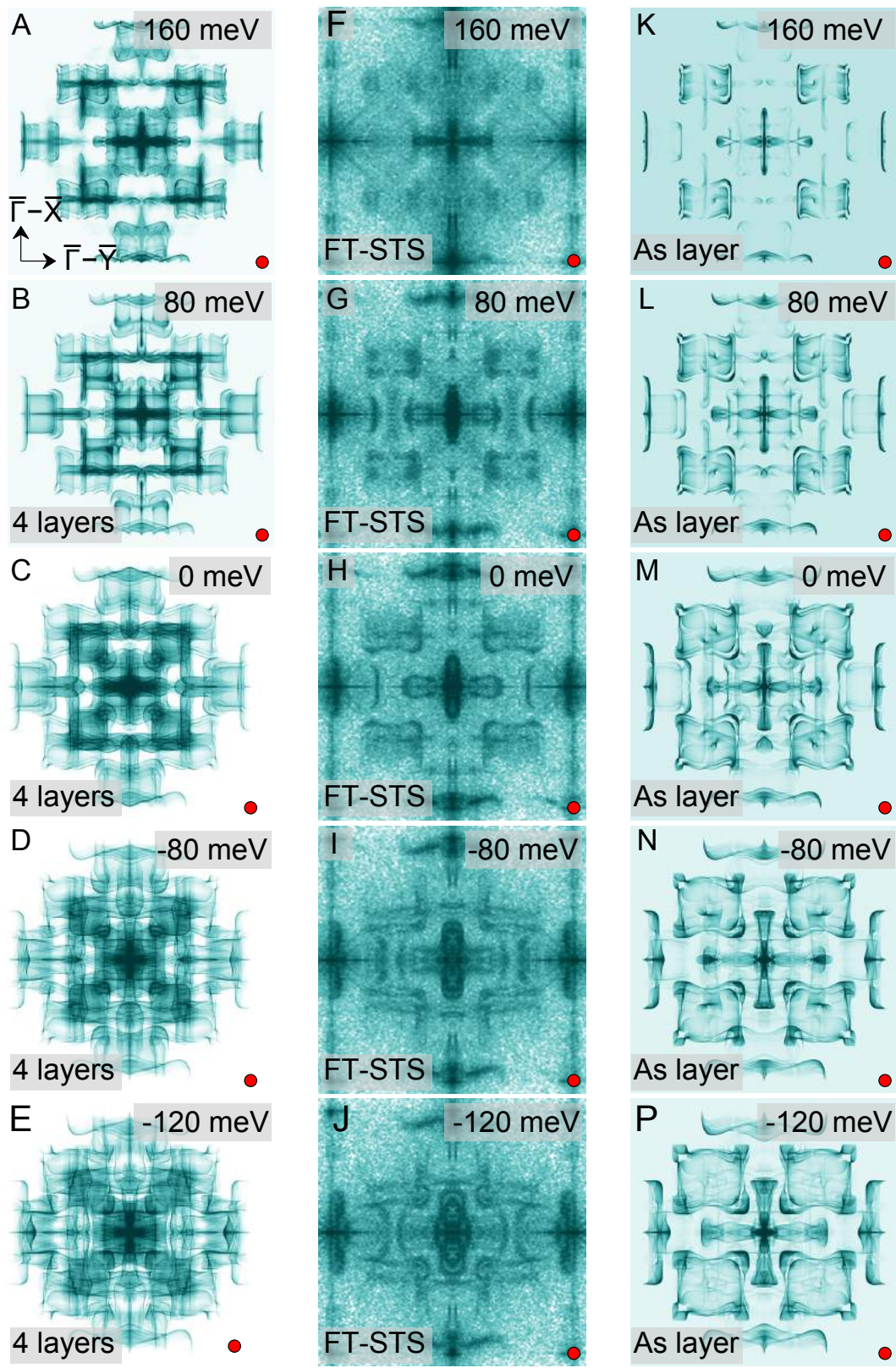

**Fig. S4.** Comparison of SSPs of the quadruple layer **(A-E)** and As layer projections **(F-J)** with the measured QPI maps **(K-P)** at various energies.

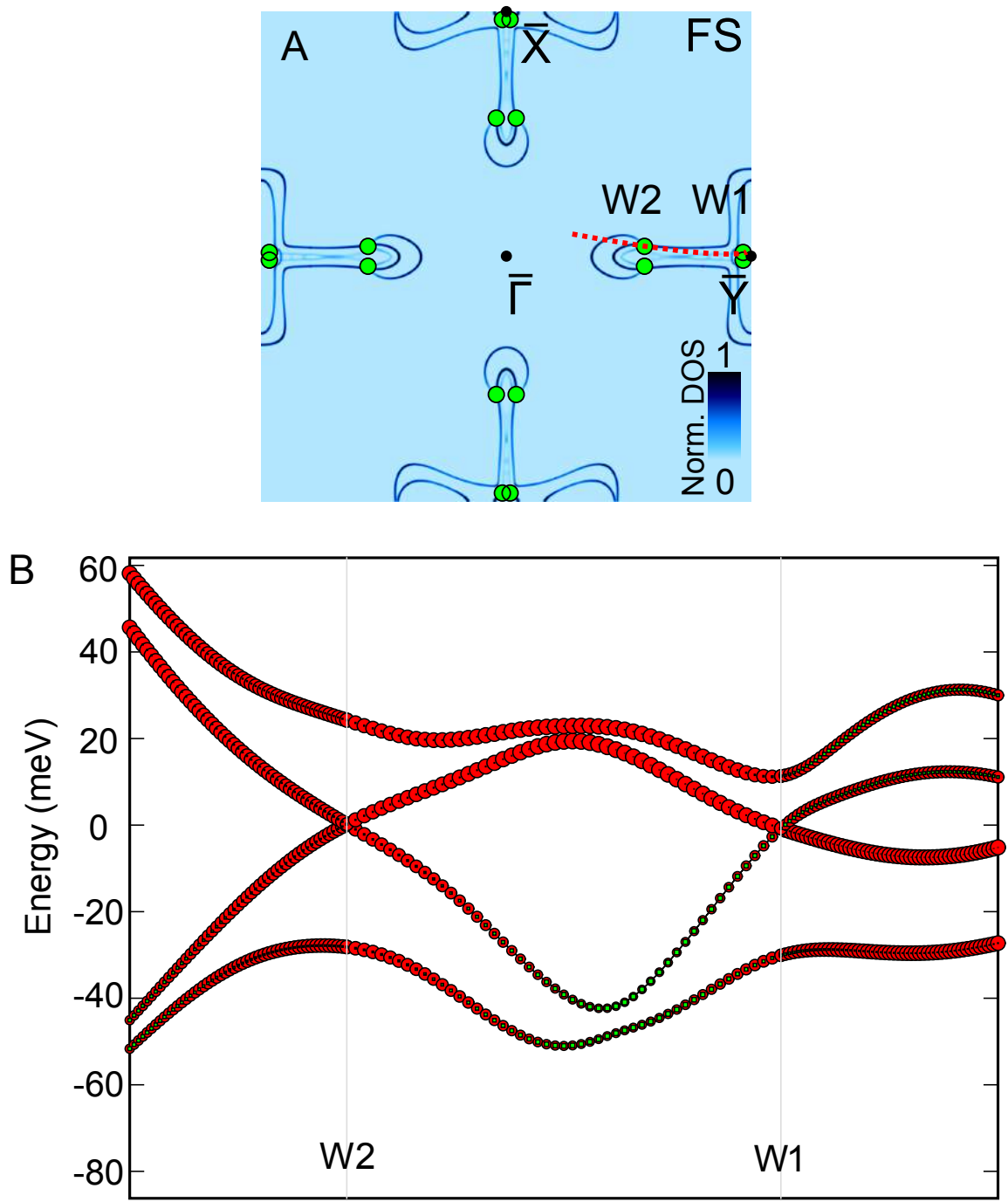

**Fig. S5.** (**A**) The projected cut and locations of Weyl points W1 and W2. (**B**) Bulk band structure close to the Fermi energy along a line of momenta crossing both types of Weyl nodes (W1 and W2). The size of the markers indicates the total weight of a state on Ta (red circle) and that on As (green square).

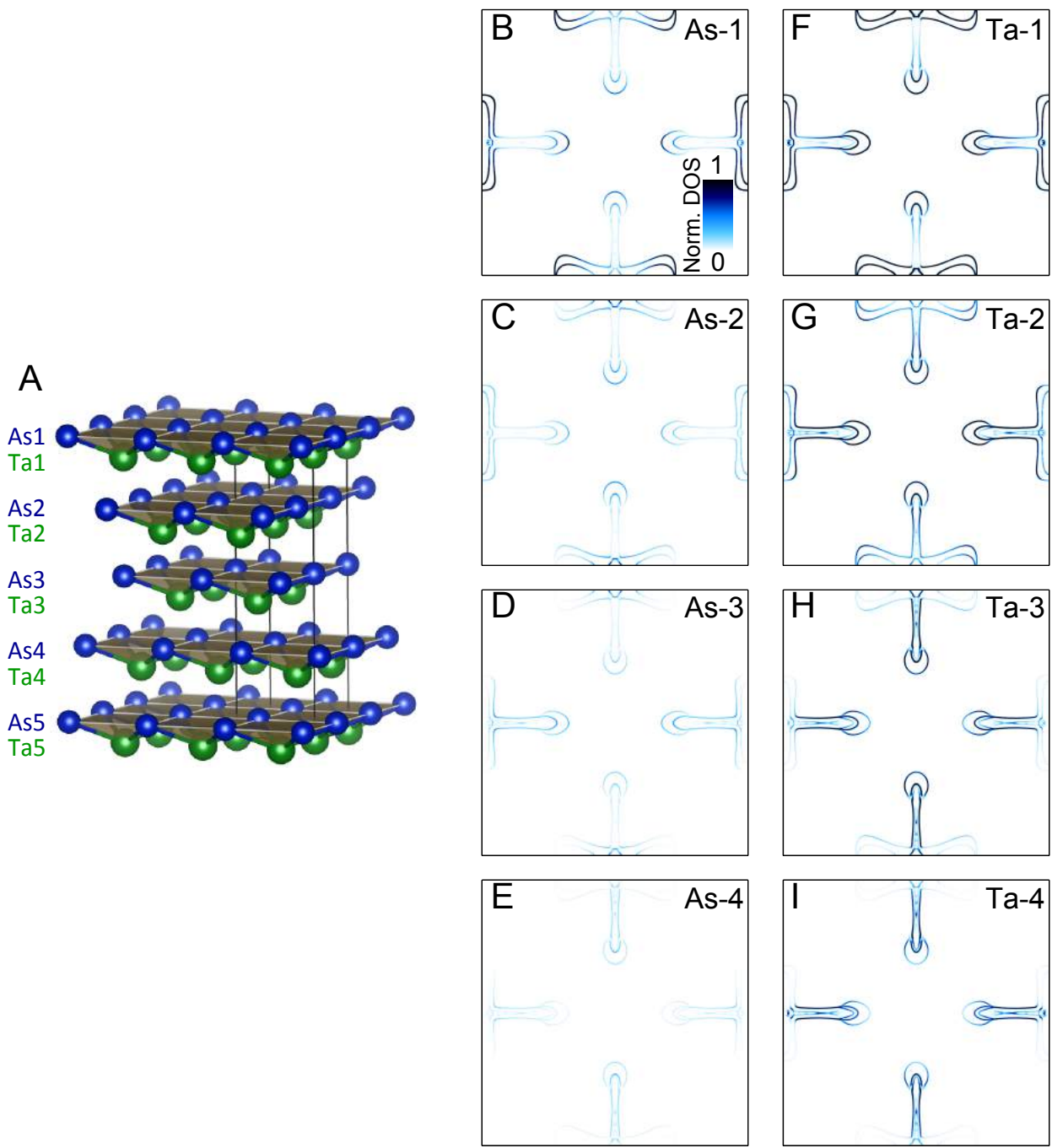

**Fig. S6.** (**A**) Schematic illustration of the TaAs crystal structure. The layer indices of As and Ta are indicated (1 stands for the topmost layer.). (**B-I**) Spectral density projected to each of As layer (**B-E**) and Ta layer (**F-I**), separately, for the top four layers (which corresponds to one unit cell).

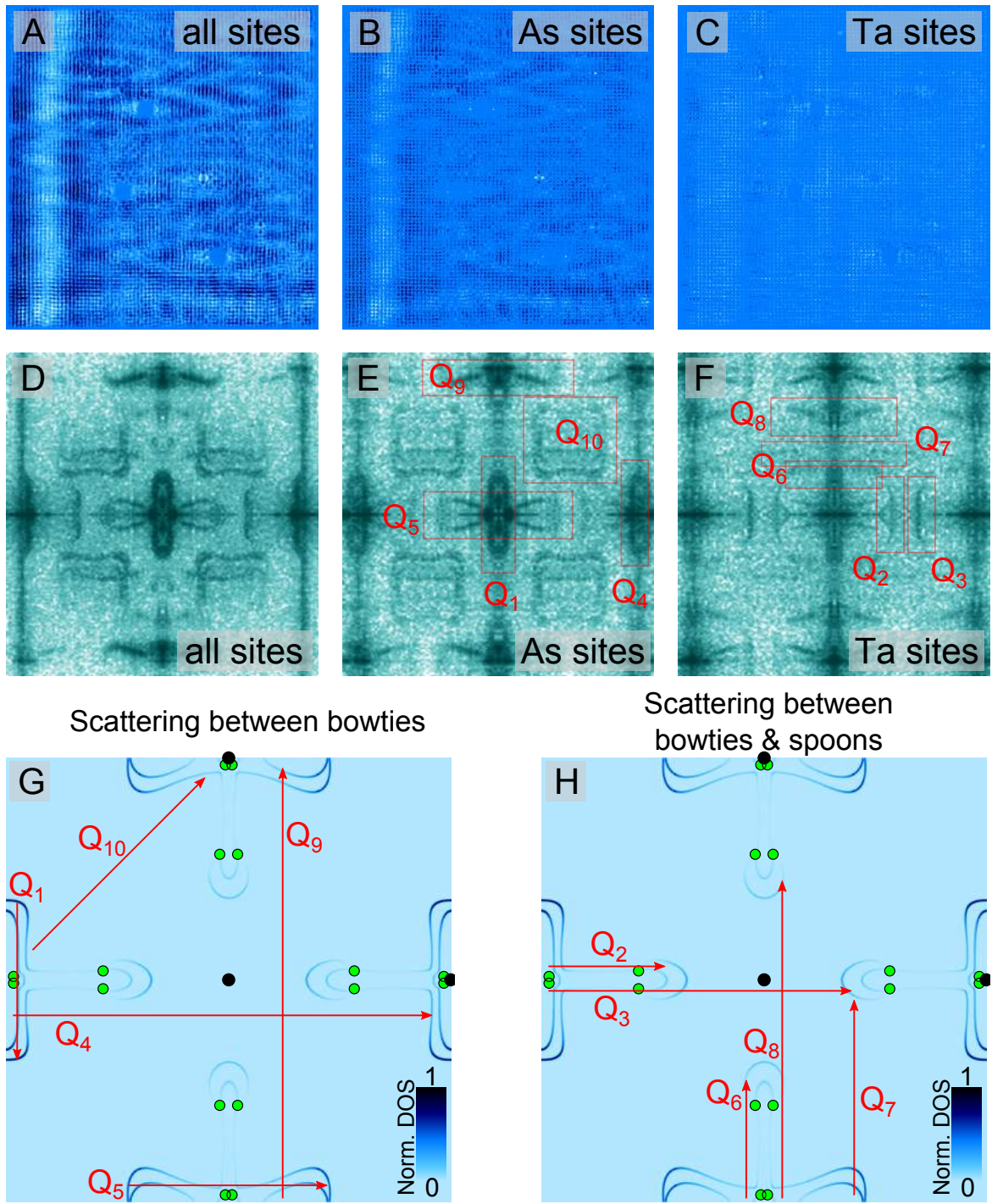

**Fig. S7.** (**A**) Real space $dI/dV$ conductance map measured on the surface of TaAs, the As-site-selected (**B**) and Ta-site-selected data (**C**) after applying the described real space masking. (**D-F**) Fourier transform of each map in (**A-C**). (**G**) The $Q$ vectors indicate the scattering wavevectors among bowties, which mainly contribute to (**E**). (**H**) The scattering wavevectors between spoons and bowties gives the main contribution of the QPI features remained in (**F**).

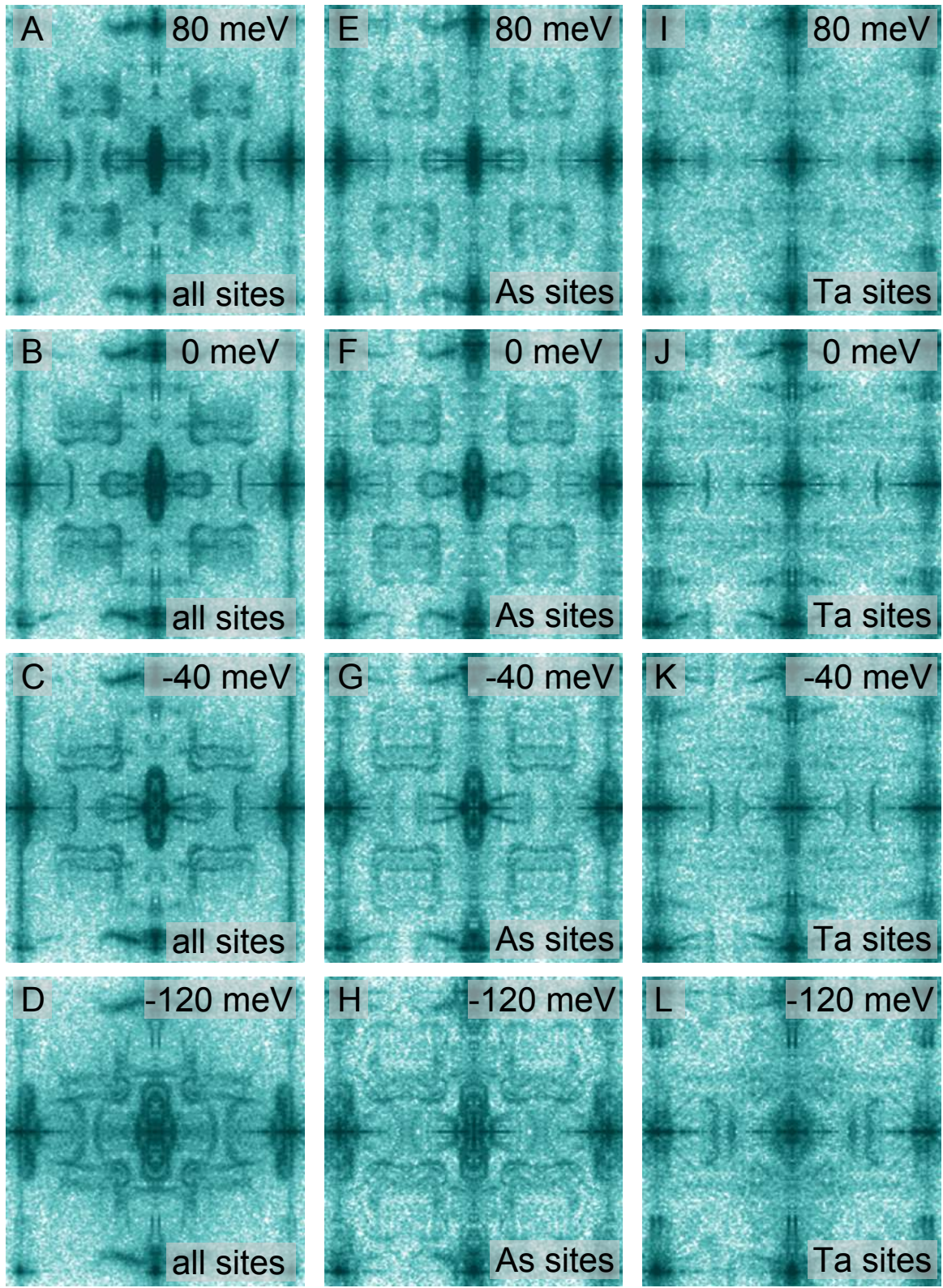

**Fig. S8.** (**A-D**) Full QPI maps acquired on the surface of TaAs at different energies. Filtered QPI maps corresponding only to the As-sites (**E-H**) and only to the Ta-sites (**I-L**) show the decomposition of QPI wavevectors at multiple energies.